\documentclass[preprint,showpacs,preprintnumbers,amsmath,amssymb]{revtex4}
\usepackage{graphicx}
\usepackage{dcolumn}
\usepackage{bm}
\begin{document}
\renewcommand{\thefootnote}{\fnsymbol{footnote}}
\title{Ultraquantum magnetoresistance in single-crystalline $\beta $-Ag$_2$Se}
\author{Chenglong Zhang$^{1}$\footnotemark[2], Haiwen Liu$^{1}$\footnotemark[2], Tay-Rong Chang$^3$,  Su-Yang Xu$^4$, Wei Hua$^5$, Hua Jiang$^6$, Zhujun Yuan$^1$, Junliang Sun$^5$, Horng-Tay Jeng$^{3,7}$, M. Zahid Hasan$^4$, X.C. Xie $^{1,2}$, Shuang Jia$^{1,2}$\footnotemark[1]}
\affiliation{$^1$ICQM, School of Physics, Peking University, Beijing 100871, China\\
$^2$Collaborative Innovation Center of Quantum Matter, Beijing, China\\
$^3$Department of Physics, National Tsing Hua University, Hsinchu 30013, Taiwan\\
$^4$Joseph Henry Laboratory, Department of Physics, Princeton University, Princeton, New Jersey 08544, USA\\
$^5$Department of Chemistry, Peking University, Beijing 100871, China\\
$^6$College of Physics, Optoelectronics and Energy, Soochow University, Suzhou 215006, China\\
$^7$Institute of Physics, Academia Sinica, Taipei 11529, Taiwan
}

\footnotetext[1]{e-mail: gwljiashuang@pku.edu.cn}
\footnotetext[2]{These authors contributed equally to this work}

\begin{abstract}
In the history of condensed matter physics, reinvestigation of a well-studied material with enhanced quality sometimes led to important scientific discoveries.
A well-known example is the discovery of fractional quantum Hall effect in high quality GaAs/AlGaAs heterojunctions.
Here we report the first single crystal growth and magnetoresistance (MR) measurements of the silver chalcogenide $\beta $-Ag$_2$Se (Naumannite), a compound has been known for the unusual, linear-field-dependent MR in its polycrystalline form for over a decade \cite{xu_large_1997, qmrprb1998abrikosov}.
With the quantum limit (QL) as low as 3 Tesla, a moderate field produced by a superconductor magnet available in many laboratories can easily drive the electrons in Ag$_2$Se to an unprecedented state.
We observed significant negative longitudinal MR beyond the QL, which was understood as a `charge-pumping' effect between the novel fermions with opposite chiralities.
Characterization of the single-crystalline Ag$_2$Se and the fabrication of electric devices working above the QL, will represent a new direction for the study of these exotic electrons.

\end{abstract}

\maketitle

Electrons in metal move in the quantized orbits known as Landau levels (LLs) in an external magnetic field \cite{mrinmetals, moinmetals}.
The so-called quantumn limit(QL) is attained when the magnetic field is higher than that corrosponds to the lowest LL, in which all the electrons are confined in one highly degenerate state.
Beyond the QL, the electrons are further fractionalized in two-dimentional(2D) electron gas, while little is known about the three-dimentional(3D) systems.
Unfortunately very few materals manifest their electron concertration dilute enough so that the QL is accessble.
Here we show that the single-crystalline $\beta $-Ag$_2$Se, a well-known quantum material, is one candidate for the exporation of this unknown region.

The mineral Naumannite ($\beta $-Ag$_2$Se) was first discovered in 1828.
It crystallizes in a non-central-symmetric structure which has an orthorhombic unit cell ($P2_12_12_1$) with two crystallographic distinct silver atoms and one selenium atom \cite{ag2seold, ag2sezaac2008} (Fig. \ref{Fig1} insets).
Previous studies showed that polycrystalline $\beta $-Ag$_2$Se manifested large, linear-field-dependent, transversal magnetoresistance (MR) at both low and high temperature regions \cite{xu_large_1997, husmann_megagauss_2002}.
This linear MR remains unsaturated as a function of magnetic field strength up to 50 Tesla (T).
It was believed that the origin of this unusual linear MR is either a quantum effect of electron correlation due to a linear dispersion at the energy where the conduction and valence bands cross \cite{qlmrep2000abrikosov, qmrprb1998abrikosov}, or a classical effect of spatial conductivity fluctuations in strongly inhomogeneous polycrystals \cite{parish_non-saturating_2003, hu_classical_2008}.
Very recently topological characteristics of electrons in Ag$_2$Se due to spin orbital coupling (SOC) were emphasized \cite{ag2te2prl2011fang}.

Although various papers have been published on the studies of polycrystalline and nano-size Ag$_2$Se \cite{ag2sepressure2014, nanoag2sejacs2001, Yu_Ag2Se}, research on a macro-size single crystal has never been reported until now.
The difficulty of the single crystal growth comes from the thermal instability of the $\beta $-phase: it undergoes a first order, polymorphic structural transition to a cubic, high-temperature $\alpha $-phase at $133 ^o$\it{C}\rm.
In the $\alpha $-phase, selenium atoms form a body-centered cubic frame and silver atoms are distributed over several interstitial sites \cite{ag2severyold, Kumashiro1996761}.
During a regular crystal growth process, $\alpha $-Ag$_2$Se intends to form a multi-domain polycrystalline $\beta $-phase when cooling to the room temperature.
By contrast, we discovered a novel modified self-selecting vapour growth technique \cite{Szczerbakow200581}, which enables us to obtain single crystals of $\beta $-Ag$_2$Se above the transition temperature (See details in Method Part and Supplemental Information (SI)).
During this growth process, the vapour from polycrystalline Ag$_2$Se condensed on small seed plates of the $\alpha $-phase, leading to the ribbon-like $\beta $-phase single crystals (Fig. \ref{Fig1} inset).
Macro-size single crystals of $\beta $-Ag$_2$Se exclusively provide a chance for studying the electronic states via transport measurements with less influence of scattering from defects, particularly from the grain boundaries in polycrystals.
We focus on the MR measurements for the current ($I$) along the crystallographic $\bf{a}$ direction while the magnetic field can be along different directions in this paper (Fig. \ref{Fig1} inset).

Temperature dependent resistivity of single crystalline Ag$_2$Se shows a similar profile as that of previously reported in  polycrystals \cite{xu_large_1997, husmann_megagauss_2002}: a broad maximum occurs around 100 K, below which the resistivity shows a metallic-like behavior (Fig. \ref{Fig1}).
The resistivity of the samples R1 and R2 that are from one growth batch shows one order of magnitude less than that of the samples R3 and R4 from the other.
Hall measurements (for $I$//$\bf{a}$, $H$//$\bf{b}$, and Hall voltage along $\bf{c}$) reveal that all samples are $n$-type for the whole temperature range and the carrier density is $n=4.9\times 10^{19}\mathrm{cm}^{-3} $ and  $2.4\times 10^{18}\mathrm{cm}^{-3} $ for the samples R1 and R4 at 2~K, respectively (See SI and Table \ref{tab}).
The $n$-type carriers most likely come from a small amount of deficiencies of selenium atoms commonly observed in selenide compounds \cite{jia_BTS_2011}.
Both R1 and R4 show mobility ($\mu _b = 1/\rho ne $) about 1000 cm$^2/(\mathrm{V\cdot s})$ at 2 K, leading to a product of the classical cyclotron frequency to the scattering time, $\omega _C\tau =\mu _bB<1$ at 9 T \cite{mrinmetals}.

Figure \ref{Fig2} shows the MR (defined as $\Delta \rho _H/\rho _0 $) of the samples R2 and R4 along the three principal axes at representative temperatures.
The transversal MR for $H$//$\bf{b}$ and $\bf{c}$ shows a similar profile: it crosses over from a quadratic dependence on $H$ at low fields, to a linear and unsaturated dependence up to 9 T.
This linear and unsaturated MR up to 300~K is similar to the previous reports in polycrystals \cite{xu_large_1997, ag2semr1999, husmann_megagauss_2002}.
For $H$//$\bf{b}$ in R1 and R2, weak Shubnikov-de Haas (SdH) oscillations occur on the linear background when $T<10$~K and $\mu _0H>5$ T, but not in R3 and R4.
Further analysis reveals that these single-frequency oscillations came from a relatively large Fermi extremal surface ($S_{F2} = 49$ T and $36$ T for R1 and R2, respectively, see more details in SI).
Given that these oscillations only appear in the samples with large carrier densities, we expect that they most likely come from a $3D$ electron band with a conventional energy dispersion.
Further measurement in higher magnetic fields is prepared.

At first glance, the longitudinal magnetoresistance (LMR) of R2 and R4 for  $H$//$\bf{a}$ shows different profiles at low temperatures: R2 shows strong SdH oscillations up to 9 T, while R4 shows weak oscillations when $H<3.2$~T.
Beyond this field corresponding to the lowest quantized LL, R4's LMR dramatically decreases with increasing $H$, \textit{ie.}, R4 shows negative LMR above the QL.
The Lifshitz-Onsager rule is employed to understand the quantization of the Fermi surface cross-section area $S_F$ as a function of the magnetic field $B$:
\begin{equation}
S_F \frac{\hbar }{eB}=2\pi (n+\gamma )
\label{eqn1}
\end{equation}
where $\hbar $ is the Planck's constant; $e$ is the elementary charge; $n$ is the LL index, and $\gamma $ is the Onsager phase.
The SdH oscillations of R2 originate from a small Fermi surface with $S_{F1}=8.1$~T, and the peaks at $\pm 6.8$~T correspond to $n=1$ (See Fig.\ref{Fig3} D and E).
It is noteworthy that when $ T=40$~K, the oscillations fade out but the negative LMR still persists.
On the other hand, the SdH oscillations in R4 originate from an even smaller Fermi surface with $S_{F1}= 3.8$~T, and a field larger than 3.2~T drives the system beyond the QL (See Fig.\ref{Fig3} F and G).
The oscillations fade out above 10~K, but the negative LMR persists until 25~K, and then it is dramatically suppressed to a very small, positive signal above 50~K.
Although R1 and R3 have different $S_F$, their LMR manifests a similar behavior in rescaled $H$ with respect to $S_F$: i), the negative LMR is always found to occur as long as the magnetic field is higher than the QL; ii), the negative LMR is not a part of oscillations since it survives at higher temperatures which destroy the oscillations.
These phenomena described above have been observed in a dozen of samples with various sizes and $n$ (see SI and Table \ref{tab} for the samples R1 and R3).
Based on the analysis we believe that the negative LMR is due to a band effect above the QL.
It is noteworthy that the LMR of R3 and R4 shows a shoulder-like anomaly around 6 T (indicated by the arrows in Fig. \ref{Fig2} and SI).
These anomalies are located at the fields that are very close to the $n=1/2$ in the LL indices (Fig. \ref{Fig4} C and E) for various samples with different $S_F$.
It is likely due to the magnetic field induced spontaneous symmetry breaking which drives the system into a fractional filled LL.
Further studies are needed to clarify this anomaly.

In order to address the Fermi surface topology and the nature of the electrons contributing to the SdH oscillations, we measured the MR while rotating the direction of the magnetic field with respect to the three principal axes at 2~K (Fig. \ref{Fig3} A, B, C and SI).
When the direction of the field changes from $\bf{a}$ to $\bf{b}$ or to $\bf{c}$, the MR of R2 shows a very similar profile with respect to the change of $\phi $ and $\psi $: the oscillations persist until $\phi $ and $\psi = 60^\circ $, while the negative MR is fully compensated by the linear contribution from the transversal MR for $\phi $ and $\psi > 30^\circ $.
Figure \ref{Fig3} D and E show the resistivity ($\rho _H$) versus the reciprocal of the field component along the $\bf{a}$ direction for R2.
The oscillations' period only depends on the field component along the $\bf{a}$ direction, whenever the field is tilted to $\bf{b}$ or $\bf{c}$.
A similar angle dependent behaviors for R4 is shown in Fig. \ref{Fig3} F, G and SI.

Based on the analysis of the SdH oscillations along different directions in R2, the frequencies are plotted as a function of the angles in Fig. \ref{Fig4} A and B.
A clear $1/cos\phi $ and $1/cos\psi $ dependence up to $60^\circ $ is observed on the Fermi surface $S_{F1}$, which indicates that the electrons are distributed highly two-dimensionally and $S_{F1}$ has the geometry of a cylinder or a highly anisotropic ellipsoid.
The LL indices for all four samples are shown in Fig.\ref{Fig4} C.
The resistivity associated with the SdH oscillations is expressed as \cite{moinmetals, rashbabook}:
\begin{equation}
\rho _H=\rho _0[1+A(B,T)cos2\pi (S_F/B+\gamma )],
\label{eqn2}
\end{equation}
where
\begin{equation}
A(B,T)\propto exp(-2\pi ^2k_BT_D/\hbar \omega _C)\cdot \frac{2\pi ^2k_BT/\hbar \omega _C}{sinh(2\pi ^2k_BT/\hbar \omega _C)}.
\label{eqn3}
\end{equation}
In Eq. \ref{eqn3}, $T_D$ is the Dingle temperature; $k_B$ is the Boltzmann's constant; and the cyclotron frequency $\omega _C = eB/m^{\star }$.
Based on Eq. \ref{eqn2}, the peak and valley positions of $ \rho _{zz}$ are indicated as integer and half-integer, respectively.
The interpolations of the lines $n$ versus $1/\mu _0H$ for all four samples show that the intercepts are clustered around $\gamma = -0.15\pm 0.05$, although $S_{F1}$ varies from 13.3~T to 3.8~T.
This near-to-zero phase shift indicates a non-trivial Berry phase associated with Dirac or Weyl fermions \cite{Mikitik_berry_2012, Murakawascience2013}.
The effective mass ($m^{\star }=eB/\omega _C $) was obtained from the temperature dependence of the peak amplitude at $n=1$ for R2 and R4, yielding 0.075 and 0.064$\mathrm{m_e}$, respectively (Fig.\ref{Fig4} D).
Other parameters including the Fermi wave vector ($k_F=\sqrt{\frac{2eS_F}{\hbar }} $), the Fermi velocity ($v_F=\hbar k_F/m^{\star }$) and the Fermi energy ($E_F=v_F^2m^{\star } $) are listed in Table \ref{tab}.
These results, together with the Hall effect measurements, are consistent with the picture in which the Fermi levels of R3 and R4 lay at lower positions than those of R1 and R2 in a linearly dispersive electron band. 

According to our measurements on various samples, we believe that the negative LMR is not a current jetting effect due to highly anisotropic or inhomogeneous conductance \cite{mrinmetals, ag2seteprl2005} or improper contacts' positions (see more details in SI), nor a weak anti-localization effect of impurity scattering which is usually observed in 2DEG in low magnetic fields \cite{weaktsui1982}.
Under the condition of $\mu _bB<1$, the classical scattering effect generally changes little with respect to the magnetic field \cite{mrinmetals}. 
The strong temperature dependence and the correlation with the QL indicate that such $>20$\% negative LMR must be cauesed by the band effect of Dirac fermions.
Recently a weaker negative LMR was observed in Bi-Sb alloy with no indication of SdH oscillations \cite{ABJbisb2013}.
Besides that, negative LMR has never been reported in other Dirac electron systems such as Bi and Cd$_3$As$_2$\cite{Tanuma01031975, Bi_quantumlimit_2009, dziawa_topological_2012,liang_ultrahigh_2014}.

The crystal structure of $\beta $-Ag$_2$Se differs from the other Dirac semimetals due to the absence of the central symmetry, which can lead to symmetry protected Weyl nodes induced by SOC \cite{Murakami_NJP_2007}. 
In order to understand the chirality of electrons in Ag$_2$Se, we analyzed the low-energy effective Hamiltonian from the space group P2$_1$2$_1$2$_1$.
In the absence of SOC, the single group of P2$_1$2$_1$2$_1$ for the spinless system possesses one-dimensional representation along $\Gamma $-X ($ \Sigma $) and $\Gamma $-Y ($\Delta$) axes in the Brillouin zone (BZ) \cite{bradley2010mathematical}.
The band structure possesses two Dirac-like bands around E$_F$ on the high-symmetric $ \Sigma $ and $\Delta$ lines, respectively (Fig. \ref{Fig5} A).
This result is consistent with the primary band structure calculation based on the density functional formalism (See SI). 
The bands hybridize each other and open gaps on the $\Sigma $ and $\Delta$ lines in the presence of SOC (Fig. \ref{Fig5} B). 
One Dirac point at a high-symmetry line ($\Sigma $ or $\Delta$) splits to two Weyl points at different quadrants of the $k_{x}-k_{y}$ plane.
The 3D Weyl Hamiltonian near $E_F$ characterizes the band dispersion and the spin texture in vicinity of the gapless point $\mathbf{k_0} = (k_{x0}, k_{y0}, 0)$ (Fig. \ref{Fig5} B).
The other Weyl point near the $\Delta$ line in this quadrant has a higher energy and stays far from E$_F$.
The two half-way screw operations of the Weyl point at $\mathbf{k}_{0}$ yield two opposite chiral Weyl points at $(-k_{x0}, k_{y0},0)$ and $(k_{x0}, -k_{y0}, 0)$, while a time reversal operation gives the Weyl point with identical chirality at $-\mathbf{k}_{0}$ (see Fig. \ref{Fig5} C).

The strong SdH oscillations with a large anisotropy in Ag$_2$Se indicate a large LL spacing when $H$//$\bf{a}$ (See SI). 
Above the QL, the electrons in a Weyl system entirely occupy the zeroth LL and only possess a momentum dispersion parallel to $H$ \cite{chiral_son_PRB2013, Ran_QH_PRB2011}.
Such a chiral anomaly pumps electrons between Weyl points with the opposite chiralities at a rate proportional to the scalar product of the electric and the magnetic field $\vec E\cdot \vec B$, giving rise to a negative LMR along the direction that connects the pair of Weyl points \cite{nielsen1983adler, AjiABJ2012}.
According to our calculations, the LLs from opposite Weyl points in Ag$_2$Se coexist around the projection point $\pm k_0$ when $H$//$\bf{a}$ (Fig. \ref{Fig5} D, see SI for details).
Above the QL, the magnetic field further gives rise to a $2\delta$ shift along the direction of the field due to a dynamical chiral shift induced by the electron-electron interactions (Fig. \ref{Fig5} D) \cite{Gorbar_Weyl_PRB2013}.
The large momentum mismatch between the Fermi points in the zeroth LL impedes the backscattering process.
Because the density of states is proportional to the magnetic field in the zeroth LL in an one dimensional momentum  space, the longitudinal magneto-conductance ($\sigma = 1/\rho _{zz} $) increases linearly with $\mu _0H$ above the QL (Fig. \ref{Fig4} E).
This linear positive magneto-conductance in Ag$_2$Se is not like an $H^2$ dependence due to the impurity scattering effect \cite{Argyres_JPCS_1959}, but more like a characteristic of anomalous transport properties for a Weyl semimetal \cite{chiral_son_PRB2013, nielsen1983adler}.

In summary, we synthesized macro-size single crystals of $\beta$-phase Ag$_2$Se and measured their resistance in magnetic fields.
While the linear transversal MR was confirmed, the LMR showed strong SdH oscillations and a large, negative component beyond the QL.
This anomalous MR was understood as an electron pumping effect induced by chiral anomaly in the zeroth LL.
Our measurements also indicated that the negative LMR will be saturated at the base temperature when the magnetic field is far beyond the QL.
Some first-step measurements hinted that the sample will become an insulator in a sufficiently high magnetic field.
Finally, we infer that the vapour transfer growth method bears more efforts for synthesizing the single crystals of new materials, especially for the thermally unstable compounds.

\begin{table}[tbp]
\caption{
Parameters in samples R1 to R4. $S_{F1}$, $k_{F1}$ and $S_{F2}$, $k_{F2}$ are determined from the SdH periods on $\bf{a}$ direction and $\bf{b}$ directions, respectively. $E_F$ and $v_F$ are obtained from the $T$ dependent amplitude of oscillations on $a$ direction. Dash entries mean quantities not measured. N/A means quantities not observed.
}
\label{tab}
\begin{ruledtabular}
\begin{tabular}{lccccccccc}
 & $n$ & $S_F1$ & $k_{F1}$ & $\gamma $ & m$^{\ast }$ & $E_F$ & $v_F$ & $S_{F2}$ & $k_{F2}$ \\
 & cm$^{-3}$ & T & {\AA}$^{-1}$ & & m$_e$ & meV & $10^5$ ms$^{-1}$ & T & {\AA}$^{-1}$ \\
\hline
R1 & $4.9\times 10^{19}$ & 13.3 & 0.02 & $-0.14\pm 0.04$ & 0.1 & 30 & 2.3 & 49 & 0.039 \\
R2 & -- & 8.1 & 0.016 & $-0.18\pm 0.07$ & 0.075 & 27 & 2.5 & 36 & 0.033\\
R3 & -- & 4.7 & 0.012 & $-0.17\pm 0.01$ & -- & -- & -- & N/A & N/A \\
R4 & $2.4\times 10^{18}$ & 3.8 & 0.011 & $-0.18\pm 0.01$ & 0.064 & 18 & 2.2 & N/A & N/A \\
\end{tabular}
\end{ruledtabular}
\end{table}

\section*{Methods}
\begin{figure}[p]
  \begin{center}
  \includegraphics[clip, width=0.9\textwidth]{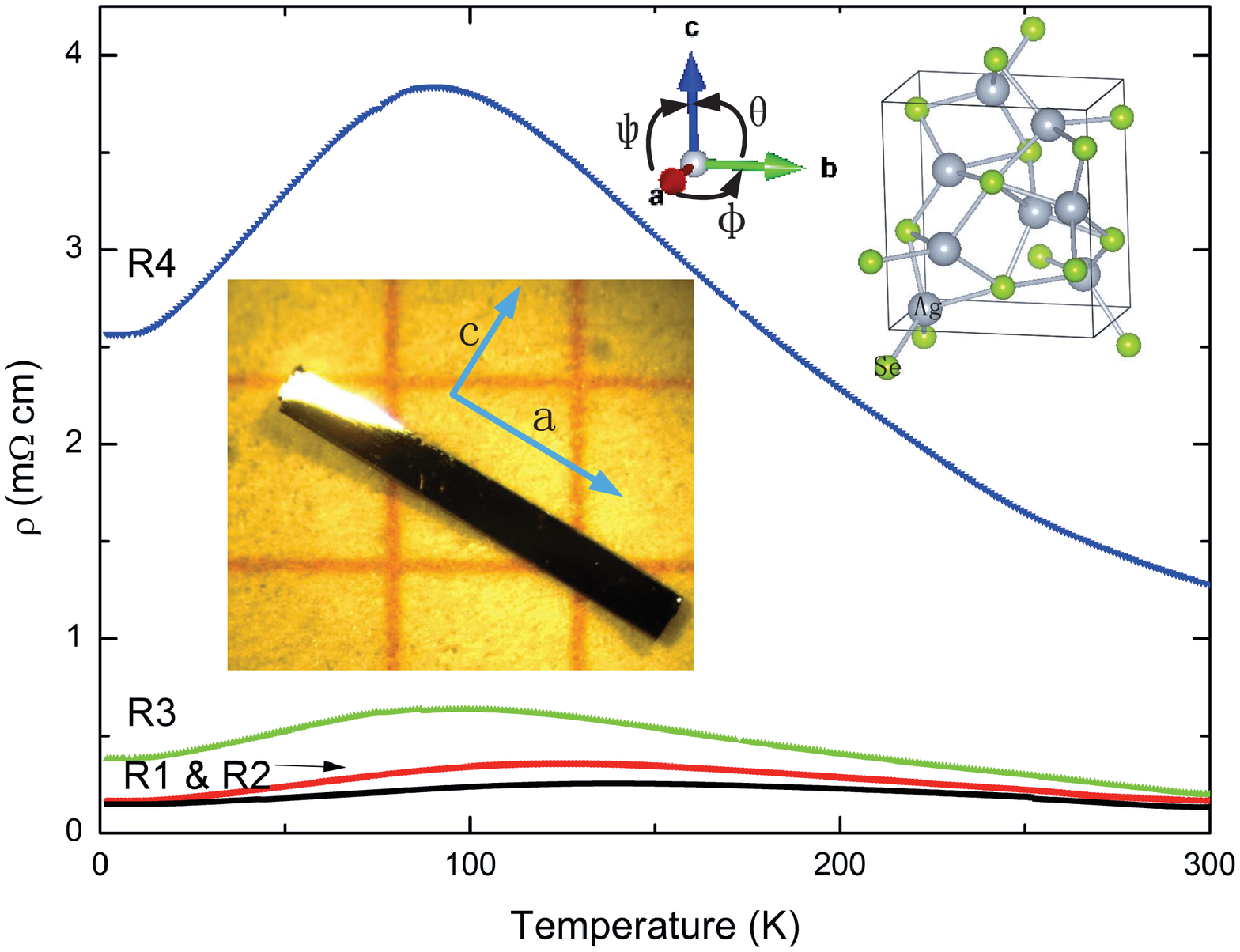}\\
  \caption{\textbf {Temperature dependent resistivity of single crystalline $\beta $-Ag$_2$Se.} The resistivity along $\bf{a}$ direction is similar as previously reported polycrystalline samples with similar carrier density $n$. Upper inset: the unit cell and the definition of angles $\theta $, $\phi $ and $\psi $ among three principal axes. Central inset: an image of a crystal under microscope. The scale is $1\times 1$ mm.}
  \label{Fig1}
  \end{center}
\end{figure}

\begin{figure}[p]
  \begin{center}
  \includegraphics[clip, width=1\textwidth]{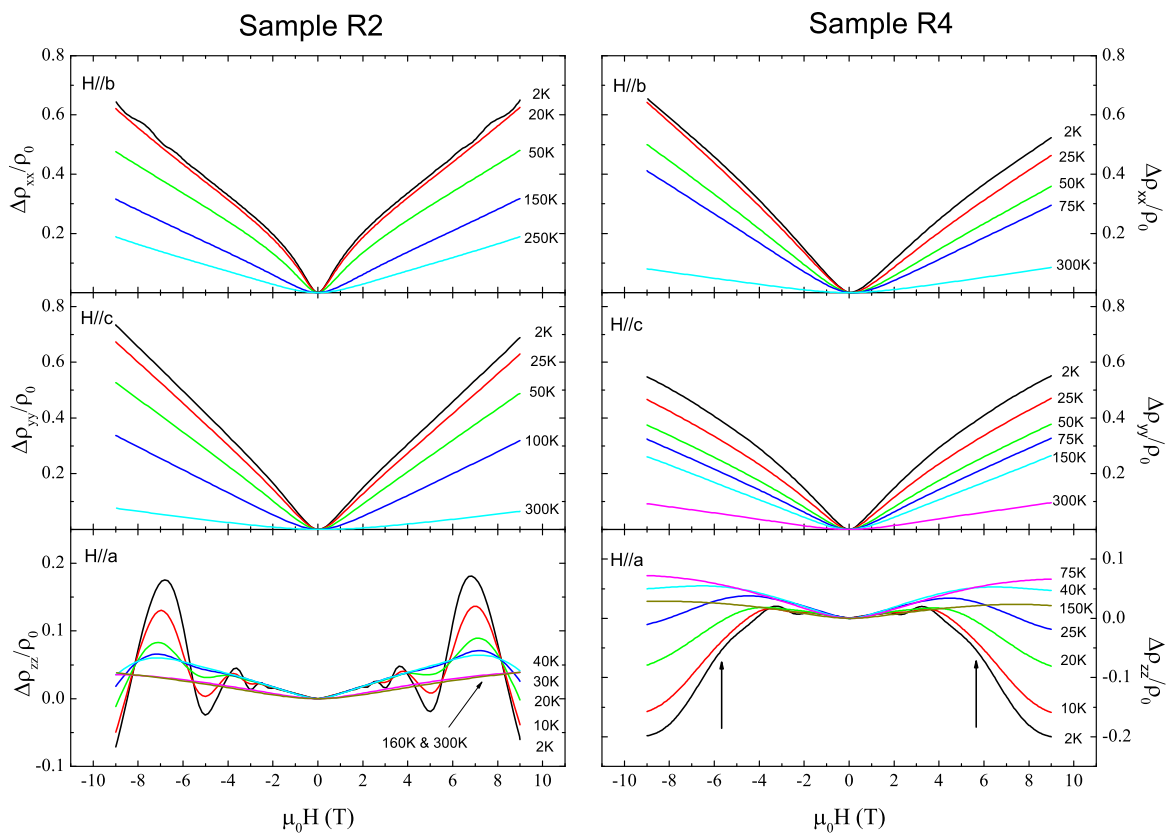}\\
  \caption{\textbf {MR ($\Delta \rho /\rho _0 $) for the samples R2 and R4 along three principal axes at representative temperatures.} $\Delta \rho _{xx}$,  $\Delta \rho _{yy}$ and  $\Delta \rho _{zz}$ are for $H$//$\bf{b}$, $\bf{c}$ and $\bf{a}$, respectively. Since the raw data are highly symmetric, the treatment of $\rho _H = (\rho _{+H}+\rho _{-H})/2$ was not applied. The arrows in the lower right panel show the shoulder-like anomaly above the QL. }
  \label{Fig2}
  \end{center}
\end{figure}

\begin{figure}[p]
  \begin{center}
  \includegraphics[clip, width=0.7\textwidth]{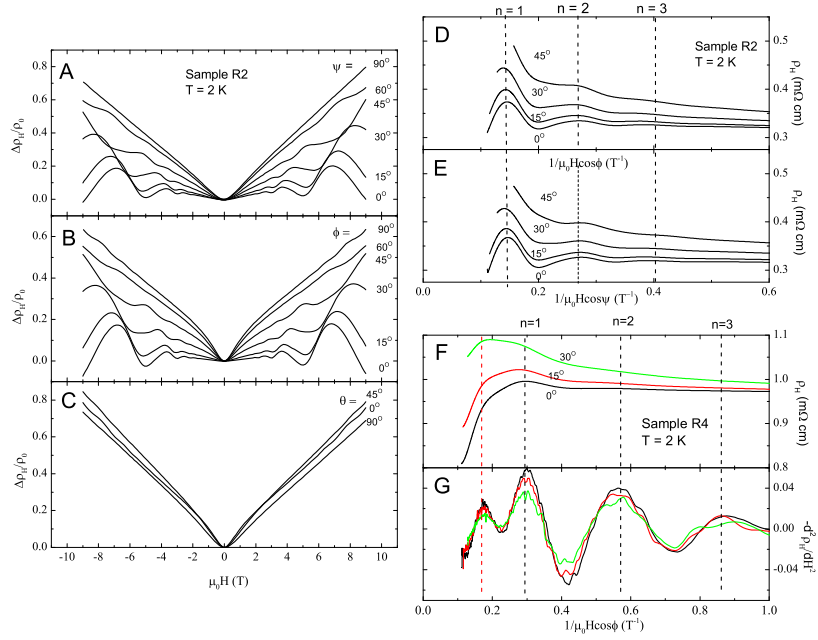}\\
 \caption{\textbf {Anisotropy of MR for the samples R2 and R4 at 2~K.} Panel A, B and C: the MR for R2 when the direction of the field change among three principal axes at 2 K.  No symmetry treatment of $\pm H$ was applied. The MR slightly changes when $\theta$ changes from $0^{\circ} $ to $90^{\circ }$. Panel D and E: the resistivity versus $1/\mu _0H_a = 1/(\mu _0Hcos\phi )$ and $1/(\mu _0Hcos\psi )$ for R2 at 2~K, respectively. Panel F and G: $\rho _H$ and $-d^2\rho _H/dH^2 $ versus $1/\mu _0H_a = 1/(\mu _0Hcos\phi )$ for R4 at 2~K. The red dashed line on the left represents the position of the shoulder-like anomaly beyond the QL (see more details in the main text).}
  \label{Fig3}
  \end{center}
\end{figure}

\begin{figure}[p]
  \begin{center}
  \includegraphics[clip, width=1\textwidth]{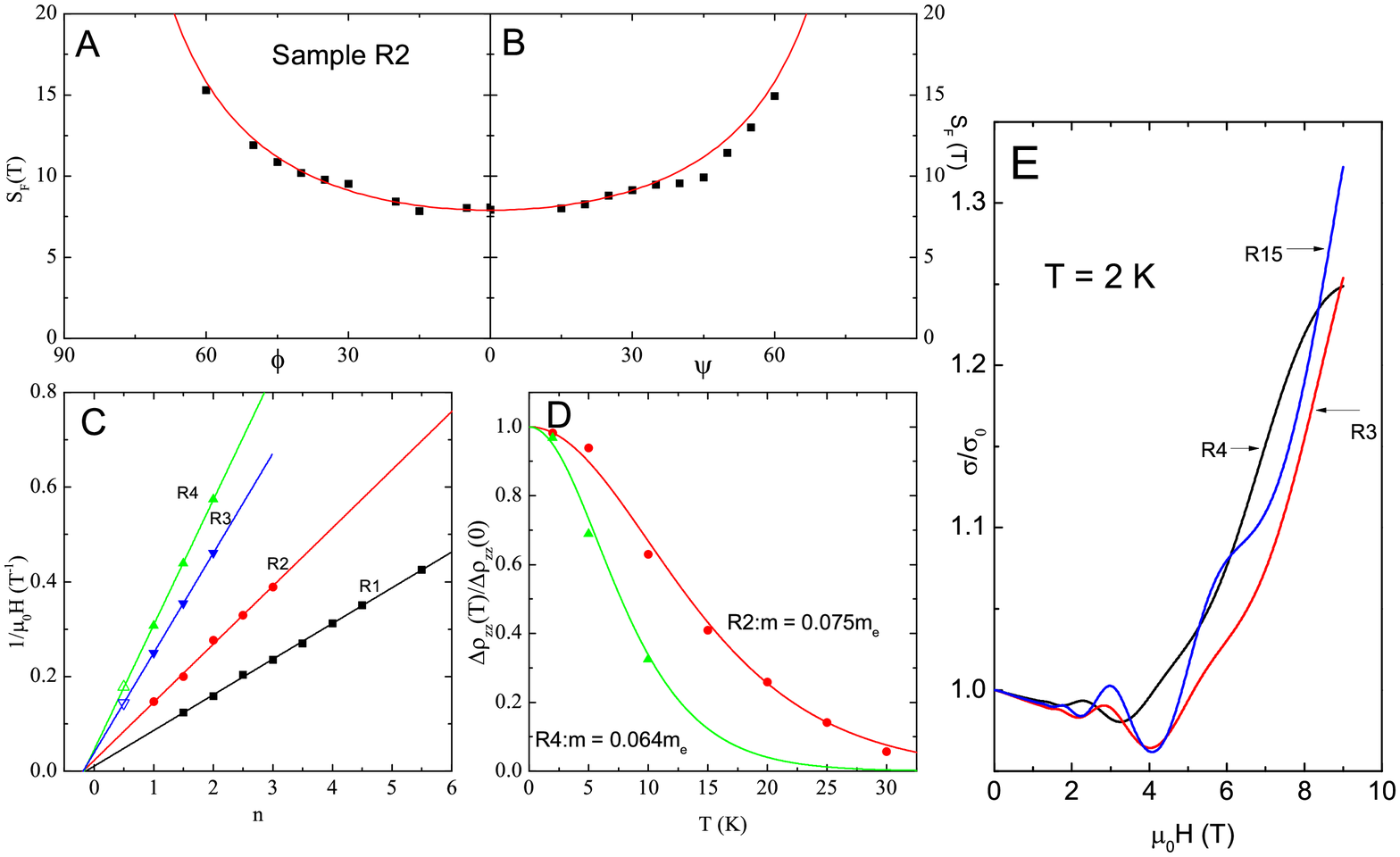}\\
  \caption{\textbf {Analysis of SdH oscillations and negative LMR in Ag$_2$Se.} Panel A and B: S$_{F1}$ for R2 varying with $\phi $ and $\psi $ as $1/cos\phi $ and $1/cos\psi $ (red curve), respectively. Panel C: Landau level indice for all 4 samples. Two open symbols present the anomaly at n = 1/2 for R3 and R4. Panel D: Temperature dependence of the SdH oscillations amplitude at n = 1 for R2 and R4. Panel E: magneto-conductance for three samples showing negative LMR below 9 T at 2~K. The arrows indicate that a linear magneto-conductance occurs when the field is higher than the position of the shoulder-like anomaly.}
  \label{Fig4}
  \end{center}
\end{figure}

\begin{figure}[p]
  \begin{center}
  \includegraphics[clip, width=1\textwidth]{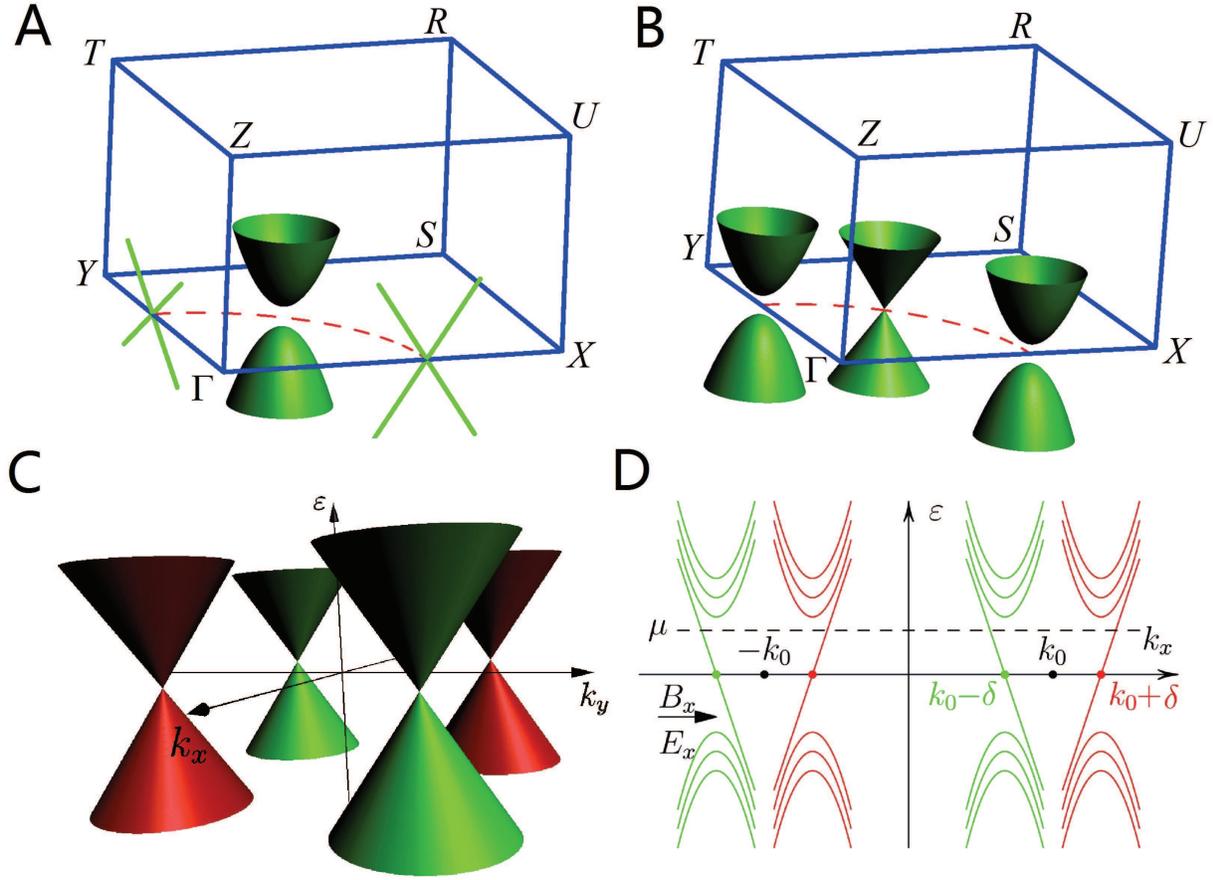}\\
  \caption{\textbf { Schematic plots of energy bands of Ag$_2$Se.} Panel A and B: The energy bands for
spinless and spinful system, respectively. Panel C: Weyl nodes in the BZ. Panel D: Landau levels for B // X when the dynamical chiral shift $2\delta $ is counted.}
  \label{Fig5}
  \end{center}
\end{figure}

Polycrystalline $\beta$-Ag$_2$Se was prepared from stoichiometric Ag and Se in a sealed quartz tube at 800 $^{\circ }C$ for two days.
The yield was then ground and sealed in a long quartz ampoule in vacuum, and placed in a tube furnace with temperature gradient from 500  $^{\circ }C$ to a temperature close to the room temperature for 7 days.
The single crystals yielded are ribbon-like and their sizes range from $20 \mu m \times 20 \mu m \times 2 mm$ to $ 0.5 mm \times 0.5 mm \times 10 mm$ (Fig. \ref{Fig1} inset).
The details of the crystal growth are discussed in SI part.
Single-crystal X-ray diffraction data was collected on a Agilent SuperNova diffractometer with a micro-focus Mo source ($\lambda = 0.7107 \AA$).
The data reduction was precessed by the CrysAlis Pro package.
Multi-scan absorption correction was applied.
All physical property characterizations were performed on a Quantum Design Physical Property Measurement System (PPMS-9) .
The first-principles calculations are based on the generalized gradient approximation (GGA) \cite{PRL_Perdew_1996} using the full-potential projected augmented wave method \cite{PRB_Blochl_1994, PRB_Kresse_1999} as implemented in the VASP package \cite{PRB_Kresse_1993, PRB_Kresse_1996, Kresse_CMS_1996}.
Electronic structure calculations were performed over $12\times 6\times 6$ Monkhorst-Pack $k$-mesh included self-consistently.

\begin{acknowledgments}
The authors thank Jun Xiong, Fang Chen and Fa Wang for helpful discussion, and Kaihui Liu for assistance with TEM measurements.
The authors thank Jian Wang and Yuan Li for using their instruments as well.
S. Jia is supported by National Basic Research Program of China (Grant Nos. 2013CB921901 and 2014CB239302).
X. C. Xie is supported by NBRPC (Grant Nos. 2012CB821402 and 2015CB921102) and NSFC under Grant No. 91221302.
J. L. Sun is supported by NSFC under Grant No. 91222107.
SYX and MZH are supported by U.S. DOE DE-FG-02-05ER46200
TRC and HTJ are supported by the National Science Council and Academia Sinica, Taiwan.
We also thank NCHC, CINC-NTU, and NCTS, Taiwan for technical support.

\end{acknowledgments}
\bibliographystyle{unsrt}
\bibliography{Ag2Se}

\end{document}